# Alkaline earth metal mediated inter-molecular magnetism in perfluorocubane dimers and chains

Zhuohang Li[1,2,†], Cong Wang[1,2,†,*], Linwei Zhou[1,2], Yurou Guan[1,2], Linlu Wu[1,2], Jiaqi Dai[1,2] and Wei Ji[1,2*]

[1]*Beijing Key Laboratory of Optoelectronic Functional Materials & Micro-Nano Devices, Department of Physics, Renmin University of China, Beijing 100872, P.R. China*
[2]*Key Laboratory of Quantum State Construction and Manipulation (Ministry of Education), Renmin University of China, Beijing 100872, China*

*Corresponding authors: wcphys@ruc.edu.cn; wji@ruc.edu.cn
† These authors contributed equally to this work.

**ABSTRACT:** Perfluorocubane ($C_8F_8$) was successfully synthesized and found to accept and store electrons in its internal cubic cavity to form magnetic moments. However their inter-molecule spin-exchange coupling mechanism is yet to be revealed. In this study, we found the inter-molecule magnetic groundstates of $C_8F_8$ dimer and one-dimensional (1D) chain are tunable from antiferromagnetic (AFM) to ferromagnetic (FM) by stacking orders and alkaline earth metals intercalation using first-principle calculations. The inter-molecule couplings are dominated by noncovalent halogen C-F…$C_4$ interactions. Stacking orders of dimers can regulate the relative position of the lone pairs and σ-holes at the molecular interface and thus the magnetic groundstates. Alkaline earth metals M (M = Na, Mg) intercalations could form $C_4$-M-$C_4$ bonds and lead to FM direct exchange at the inter-molecule region. An unpaired electron donated by the intercalated atoms or electron doping can result in a local magnetic moment in dimers, exhibiting an on−off switching by the odd−even number of electron filling. Novel electronic properties such as spin gapless semiconductor and charge density wave (CDW) states emerge when $C_8F_8$ molecules self-assemble with intercalated atoms to form 1D chains. These findings manifest the roles of stacking and intercalation in modifying intermolecular magnetism and the revealed halogen bond-dominated exchange mechanisms are paramount additions to those previously established non-covalent couplings.

## I. INTRODUCTION

Clusters or molecules with precise structures that exhibit collective behaviors akin to traditional atoms, represent a frontier in material science [1–3]. Magnetic clusters have received increasing attention due to their potential applications in spintronics, quantum computing and information storage [4]. Previous researches usually consider introducing metal atoms to construct magnetic clusters, including metal clusters and embedded metal clusters, e.g., $Rh_{19}^-$ [5], $M_{13}O_8$ [6]，$Cr@Si_{12}$ [7,8], $Ti@Au_{14}$ [7], $U@B_{40}$ [7] and $Mn@Sn_{12}$ [9]. In these systems, metal atoms could introduce unpaired electrons for local magnetic moments and participate in magnetic coupling within and between clusters. Compared to metal-containing clusters, the magnetic coupling mechanism of magnetic clusters composed entirely of nonmetal atoms has received relatively little attention. Electron-deficient boron clusters [10,11]and charged clusters [12,13] are two-types of metal-free magnetic clusters, of which the magnetic coupling mechanism remains to be revealed.

Magnetic clusters can be assembled as superatoms to form dimers [7–10], one-dimensional (1D) chains [14,15]，two-dimensional (2D) [16,17] and three-dimensional (3D) crystals [18,19], showing excellent physical properties and tunability, which have attracted wide attention. Endogenous metal-doped clusters such as $Cr@Si_{12}$ and $Mn@Sn_{12}$ can be assembled into dimers through covalent bonds [7–9]. Furthermore, density functional theory (DFT) calculations predict FM/ferrimagnetic (FiM)/AFM fullerene-based $U_2C@C_{80}$-M (M = Cr, Mn, Mo, and Ru) one-dimensional chains [14], and an one-dimensional AFM chain comprised of AgCu nanoclusters bonded through sulfur atoms [15]. $B_{63}$ clusters can form a stable bilayer structure through covalent B-B bonds [10]. However, there is currently limited research on the magnetic properties of magnetic clusters composed of non-metal atoms forming dimers or periodic structures through non-covalent interactions.

Perfluorocubane ($C_8F_8$), which was proposed in 2004 [20], can store an electron in its cage cavity and form magnetic moments [21–24]. Experiments in 2022 reported the successful synthesis of $C_8F_8$, confirming its ability to store electrons, while the binding between the clusters was considered to be a non-covalent interaction [25]. $C_8F_8$ provides us with the smallest cluster unit with the simplest structure among magnetic clusters and charged clusters. However, the inter-molecule magnetic coupling mechanism and potential modulating approach are yet to be revealed.

Here, we carried out fist-principle calculations to unveil the stacking order and Alkaline earth metal intercalation dependent inter-molecule electronic and magnetic coulings in the $[C_8F_8]^-$ dimer and 1D chains. We found the inter-molecule couplings are dominated by noncovalent halogen C-F···$C_4$ interactions. Stacking orders of dimers can regulate the relative position of the lone pairs and σ-holes at the molecular interface and thus the magnetic groundstates. In addition, we have achieved the modulation of the magnetic ground state by doping alkaline earth metals, which is caused by the formation of $C_4$-M-$C_4$ bonds and FM direct exchange at the inter-molecule region. In addition, we also considered the effect of doping levels on the magnetic properties of dimer and 1D chains. Spin gapless semiconductor and charge density wave (CDW) states were found in Na-doped $C_8F_8$ chain and Mg doped $[C_8F_8]^-$ chain, respectively.

## II. COMPUTATIONAL METHODS

Our density functional theory (DFT) calculations were carried out using the generalized gradient approximation and the projector augmented wave method [26,27] as implemented in the Vienna ab-initio simulation package (VASP) [28]. Only the Gamma point was used to sample the Brillouin zone for all superatom monomer and dimer calculations. For one-dimensional molecular chains 8 × 1 × 1 K-mesh was used. A kinetic energy cutoff of 600 eV for the plane-wave basis set was used for structural relaxations and total energy calculations, which ensures the convergence of relative energies better than 0.1 meV/ $C_8F_8$. A sufficiently large vacuum layer of over 15 Å was adopted to eliminate imaging interactions with adjacent periods. All atoms were fully relaxed until the residual force per atom was less than 0.01 eV/Å in calculations at the equilibrium distance for the dimer. Dispersion correction was made at the zero-damped PBE-D3 level [29], which was proved to be accurate in describing structural-related properties of [30] and was adopted for many previous calculations [7,30,31]. Charge doping on C atoms was realized with the ionic potential method. For electron doping, electrons were removed from a 2s core level of C and placed into the lowest unoccupied band, adding 1/8e of each C in $C_8F_8$+ 1e. This method ensured that the doped charges were located in the cage. It also retained the neutrality of the whole supercell without introducing background charge, which eliminated the effects of compensating charges. The differential charge density was derived by $\Delta\rho_d = \rho_0 - \rho_1 - \rho_2 - \rho_m$, where $\rho_0$ is the charge density of dimers or primitive cell in chains, $\rho_1$ and $\rho_2$ are those of two perfluorocubanes, respectively, $\rho_m$ is the charge density of doped metal. The spin density is derived by subtracting the charge densities of the two spin components.

### III. RESULTS AND DISCUSSION

As shown in Fig. 1a, $C_8F_8$ is formed by substituting F for H in cubane. The carbon-fluorine (C-F) antibonding (σ*) orbitals overlap inside the cage, giving the $C_8F_8$ molecule the ability to store electrons inside the cage [24]. We thus explored the effect of electron doping on the magnetic properties of a single $C_8F_8$ molecule. Neutral $C_8F_8$ exhibit non-magnetic (NM) order with zero net and local magnetic moments. Under electron doping level of 1 $e$ per $C_8F_8$ ($[C_8F_8]^-$), intra-molecule ferromagnetic (FM) order becomes magnetic ground states and is at least 65.4 meV more stable than the NM order. Both C (0.012 μB) and F (0.043μB) have local magnetic moments and a single $[C_8F_8]^-$ molecule has a net magnetic moment of 1 μB. We considered two different intra-molecule anti-ferromagnetic (AFM) orders and found them cannot stably hold their initial magnetic configurations during the relaxation process and will transform to the NM order. The C-F and C-C bond lengths of a [$C_8F_8$]⁻ molecule are 1.37 Å and 1.55 Å, respectively, which are in good agreement with literature[31]. Fig. 1e shows the perspective view of the atomic differential charge density of a [$C_8F_8$]⁻ in the FM configuration. Here we define the C-F bond direction as the *z* axis and the plane perpendicular to the C-F bond as the *xy*-plane. Significant charge reduction can be found around the cyclobutane ring center and at F-$p_z$ orbitals, creating σ-holes (marked with red rectangle and circles in Fig.1e, respectively). Charge accumulation reside mainly at C-C bond region and the F-$p_{x/y}$ orbitals, forming lone pairs on F atoms (marked with black rectangle in Fig. 2e) and suggesting hybridization and thus the direct FM exchange between C and F atoms. As shown in Fig. 1f, the σ-holes (F $p_{x/y}$) are spin polarized in the opposite direction to the lone pairs(F $p_z$), .similar to that of I atoms in CrI₃ [32,33], indicating that there may also be a magnetic ground state regulated by stacking configurations between $C_8F_8$ molecules

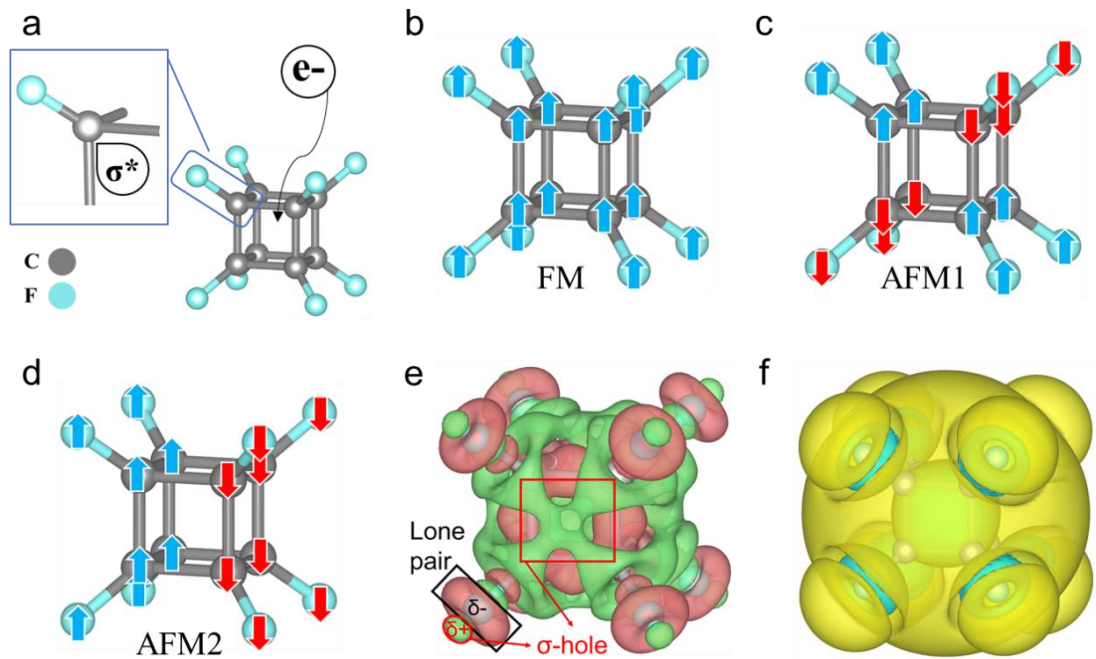

**Fig. 1 Schematic model and magnetic properties of a single [$C_8F_8$]⁻ molecule.** (a) Schematic model of a $C_8F_8$ molecule. The silver gray and cyan balls represent C and F atoms, respectively. (b-d) schematic representation of considered intra-molecule magnetic orders of a [$C_8F_8$]⁻ molecule. The blue and red arrows represent the magnetic moments up and down, respectively. (e) Atomic differential charge density of a FM $[C_8F_8]^-$ molecule. Here red and green isosurface contours correspond to charge accumulation and reduction after C-F bonding together. The isosurface value was set to 0.01 $e$/Bohr$^3$. (f) Spin density of a FM $[C_8F_8]^-$ molecule. Here yellow and blue isosurface contours correspond to spin-up and electron densities, respectively. The isosurface value were set to 7.8e-5 $e$/Bohr$^3$.

To explore the inter-molecule magnetic coupling mechanism, we further calculated the stacking configurations and magnetic groundstate of a $[C_8F_8]^-$ dimer. Here we define the cyclobutane rings (C$_4$) center, C top, C-C bonds and C-F bonds as face, top, bond and corner, respectively. To cover the possibility of stacking orders as complete as possible, we chose a high symmetry stacking order, i.e., **d11** with face to face stacking order (Fig. 2b), as initial configurations(labeled **1** in Fig. 2a). Based on this configuration, four kinds of stacking configurations were constructed by considering four kinds of 45-degree rotation operations along different axes (labeled **2** through **5** in Fig. 2a). We further considered sliding perpendicular to the direction of inter-molecular connections and combination of the above operations. Taking into account the equivalent configurations resulting from the structural symmetry of the C$_8$F$_8$ molecule (O$_h$), the calculated configurations are eventually reduced to 11 kinds of dimers. Fig. 2e plots the binding energy - intermolecule distance relation of the 11 considered stacking orders. Detailed geometric and binding energies are shown in Supplementary Figure S1,2,3 and Table S1. Intermolecule distance $d_{F-C4}$ varies from 2.56 to 6.80 Å and the energy differs by up to 0.236 eV with a nearly positive dependence on the interlayer distance. This dependence could be a result of the subtle balance between intermolecule vdW attractions and Pauli/Coulomb repulsions. For example, intermolecule F atoms exactly face the F atoms in the other $[C_8F_8]^-$ in the **d11** stacking, leading to the strongest repulsion between F atoms and thus the largest $d_{F-C4}$ of 6.80 Å and highest positive binding energy of 0.034 eV. Among stacking configurations constructed by rotation operation, **d13** with a C-F bond of one $[C_8F_8]^-$ molecule pointing to the C$_4$ face center of the other molecule (corner to face, Fig. 2c) is energetically favored with an energy gain of at least 3.6 meV. Subsequent sliding of [C$_8$F$_8$]$^-$ can stagger inter-molecule F atoms away from each other and point to the C$_4$ center of the other molecule (**d12-c**, **d14-c**, **d15-c**, and **d11-c,** Supplementary Figure S3) , leading to significantly lowered energies of at least 32meV (Fig. 2e). Configurations with intermolecule F atoms pointing to C-top (corner-to-top) and C-C bonds (corner-to-bond) sites of adjacent molecule were also calculated and are meta-stable with 0.04-0.07 eV higher binding energies compared to those corner-to-face configurations (Supplementary Table S2, S3). The AFM **d11-c** dimer is found to be the most stable stacking configuration and is at least 20 meV more stable compared with other configurations.

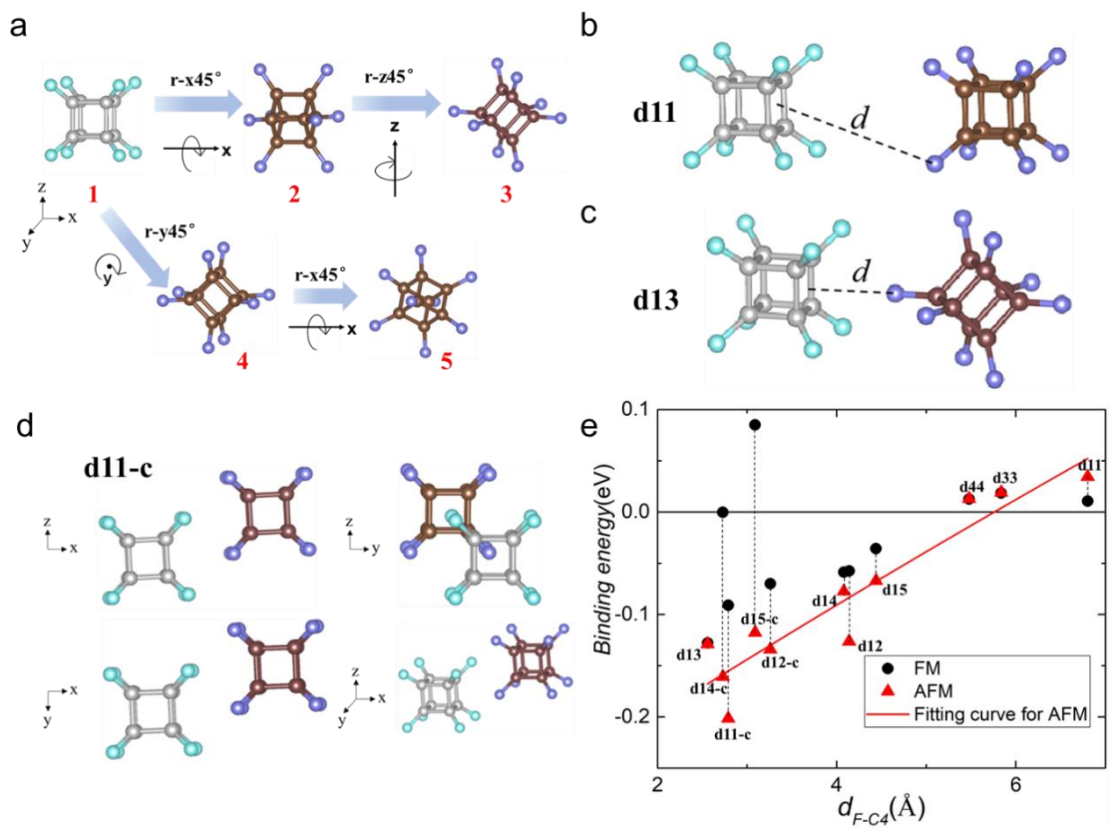

**Fig. 2 Inter-molecule stacking orders of [C$_8$F$_8$]$^-$ dimer.** (a) Five rotation operations considered in constructing [C$_8$F$_8$]$^-$ dimers. (bc) Perspective views of **d11** and **d13** configurations. (d) Top, side and perspective views of the most stable d11-c configuration. (e) Binding energies of considered dimer

configurations as a function of intermelecule distance $d_{F\text{-}C_4}$, Here the $d_{F\text{-}C_4}$ terms the nearest distance between F atom of one molecule and the center of $C_4$ ring of the other molecule, marked with dashed lines in (b-d). Black and red points correspond to binding energies of FM and AFM coupled dimers, respectively.

In addition to the thermodynamic stability, the magnetic groundstate of dimers also exhibits a stacking order dependence. Dimers with negative binding energies, such as the most stable **d11-c**, all show intermolecular AFM ground states. The inter-molecule magnetic groundstate becomes ferromagnetic when the stacking configuration is changed to **d11**, **d33** and **d44** with positive binding energies (Fig. 2e). Given the established magnetic groundstate, we further examined the stacking difference in the $[C_8F_8]^-$ dimer. The above three FM dimers share a similar structural characteristic of inter-molecule F atoms pointing exactly toward F atoms of the other molecule. While in the AFM coupled dimers such as **d11-c**, the inter-molecular F atoms are directly opposite the $C_4$ ring of the other molecule, indicating the stacking order tunable magnetic groundstate may come from the interactions between F atoms and $C_4$ rings at the inter-molecule region.

Fig.3a shows the atomic DCD of the **d11-c** stacked $[C_8F_8]^-$ dimer. The lone pair of the intermolecular F atom is closest to the σ-hole of the $C_4$ ring of the neighboring molecule, forming an intermolecular C-F…$C_4$ halogen bond and dominating the intermolecular magnetic coupling. The bond angle of the C-F…$C_4$ is 124.8°(Supplementary Figure S4), which is similar to that of typical halogen bonds [34–37]. We further plotted the inter-molecular DCD after two $[C_8F_8]^-$ molecules stacking in the **d11** manner (Fig. 3b). Significant charge transfer can be observed from the C-F bond to the lone pair of the inter-molecule F atom, leading to the increasing polarity of the C-F bond. In the molecular orbital energy level of the dimer (Fig. 3d), the C-F bonding state is labeled as HUMO-BS (Fig. 3e). The C-F bond energy is 1.35 eV lower than that of a single molecule(Supplementary Figure S5), which, together with the intermolecular charge transfer above, further confirms the picture of intermolecular halogen bond formation. The inter-molecule halogen bond results in a direct ferromagnetic coupling between σ-hole of the $C_4$ ring and the lone pair of F in a neighboring molecule. The lone pair of F within a single molecule is opposite to its σ-hole and even to the net spin polarization of the whole $[C_8F_8]^-$ molecule, leading to the inter-molecule AFM coupling in the **d11-c** stacked dimer. We further considered the influence of intermolecular distance on the magnetic ground state with constraint **d11-c** stacking order (Supplementary Figure S6). With the increasing intermolecular distance, the intermolecule AFM coupling would gradually weaken but would not be transformed into FM. Therefore, the possible influence of the inter-molecular distance between different stacking layers on the magnetic ground state was ruled out.

We further considered the case of $[C_8F_8]^-$ molecules forming one-dimensional chains. Our calculations show that the molecules are also connected by C-F…$C_4$ halogen bonds (Fig. 3g), resulting in the formation of one-dimensional Neel AFM chains (Fig. 3h).

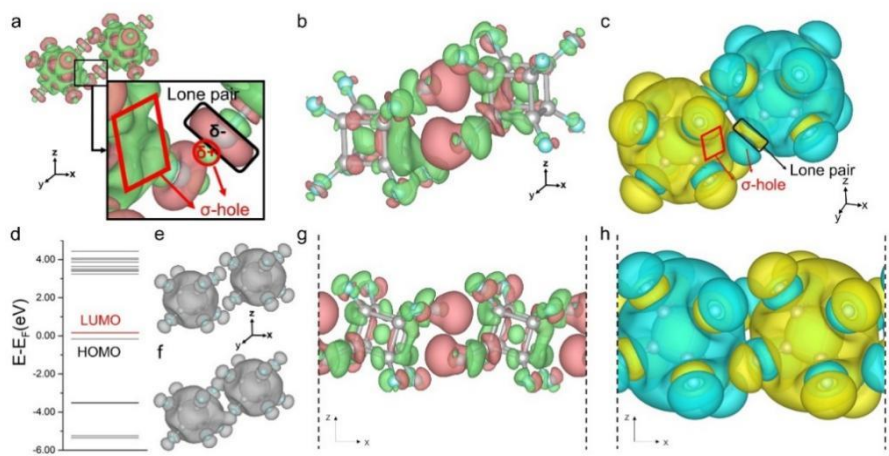

**Fig. 3 Magnetic coupling mechanism of $[C_8F_8]^-$ dimers and one-dimensional chains.** (a)Atomic differential charge density of **d11-c** stacked $[C_8F_8]^-$ dimers. Here red and green isosurface contours correspond to charge accumulation and reduction after C-F bonding together. The isosurface value was set to 0.01 $e$/Bohr$^3$. (b) Inter-molecular differential charge density of $[C_8F_8]^-$ dimers. The isosurface value was set to 0.0003 $e$/Bohr$^3$. (c) Spin density of $[C_8F_8]^-$ dimers. Here yellow and blue isosurface contours correspond to spin-up and down electron densities, respectively. The isosurface value were set to 1.5e-5 $e$/Bohr$^3$. (d) Schematic of molecular orbital energy levels, with Fermi

energy level taken as the middle of lowest unoccupied molecular orbital (LUMO) and highest occupied molecular orbital (HOMO). (e-f) The corresponding wave functions norms for LUMO (e) and HOMO (f), respectively. The isosurface value were set to 1.5e-5 $e$/Bohr$^3$. (g) Inter-molecular differential charge density of [$C_8F_8$]$^-$ 1D chain. An isosurface value of 0.0003 $e$/Bohr$^3$ was used. (h) Spin density of [$C_8F_8$]$^-$ 1D chain. The isosurface value were set to 3e-5 $e$/Bohr$^3$.

Although stacking order has been proved to be effective in tuning the magnetic coupling between [$C_8F_8$]$^-$ molecules, the positive binding energies of FM coupled stacking structures indicate their thermodynamic instability and hinder further modulation. Atomic intercalation molecules/clusters can trigger changes in inter-molecule interactions and thus modifying properties of parents materials [38]. We thus further considered alkaline earth metal Na and Mg intercalation on $C_8F_8$ dimers and 1D chains. We constructed a Na doped [$C_8F_8$]$^-$ dimer configuration by intercalating a Na atom between **d11-c** stacked [$C_8F_8$]$^-$ molecules. [$C_8F_8$]$^-$ molecules can not stably hold their position and would transform into **d11** stacking order with inter-molecule FM groundstates. The binding energy of the Na-doped [$C_8F_8$]$^-$ dimer is -1.42 eV and is one order of magnitude larger than that of the undoped **d11-c** stacking (-0.20 eV), indicating stronger ionic bonds formed between Na and [$C_8F_8$]$^-$ after intercalation.

Fig. 4a depicts the intermolecular DCD of the Na-doped [$C_8F_8$]$^-$ dimer and charge transfer from Na atoms to adjacent $C_4$ σ-hole region can be observed. Combined with the spin density (Fig. 4b), we can find that the C-F…$C_4$ halogen bonds are no longer present in the Na-doped [$C_8F_8$]$^-$ dimer, and the $C_4$-Na-$C_4$ interaction dominates the intermolecule exchange, resulting in the magnetic ground state being FM. Four midgap states are introduced by electron doping, labeled as HOMO-1, HOMO, SOMO and SUMO in Fig.4c. HOMO-1 and HOMO correspond to intermolecular bonding states mediated by intercalated atoms (Fig. 4f and Supplementary Figure S7), while SOMO-AS1 and SUMO-AS2 correspond to intermolecular antibonding states (Fig. 4d-e). Na doping brings an extra electron to fill the SOMO-AS1 orbital, resulting in spin splitting and the generation of a net magnetic moment.

The existence of the magnetic moment relies on the electron-filling number of the four midgap states, namely, whether there is an unpaired spin. We further explored the effect of different doping levels on the magnetic properties of the system: Na-doped $C_8F_8$ dimer (1 $e$ per dimer) (Supplementary Figure S8), Mg-doped $C_8F_8$ dimer (2 $e$ per dimer), Na-doped $[C_8F_8]^-$ dimer (3 $e$ per dimer) and Mg-doped $[C_8F_8]^-$ dimer (4 $e$ per dimer). Figure 4g shows the odd-even oscillation of magnetic moment and the energy difference $E_{AS1}$-$E_{AS2}$ as a function of doping level, indicating atomic intercalation to be a promising approach manipulating the on/off of magnetic moment of $C_8F_8$ dimer.

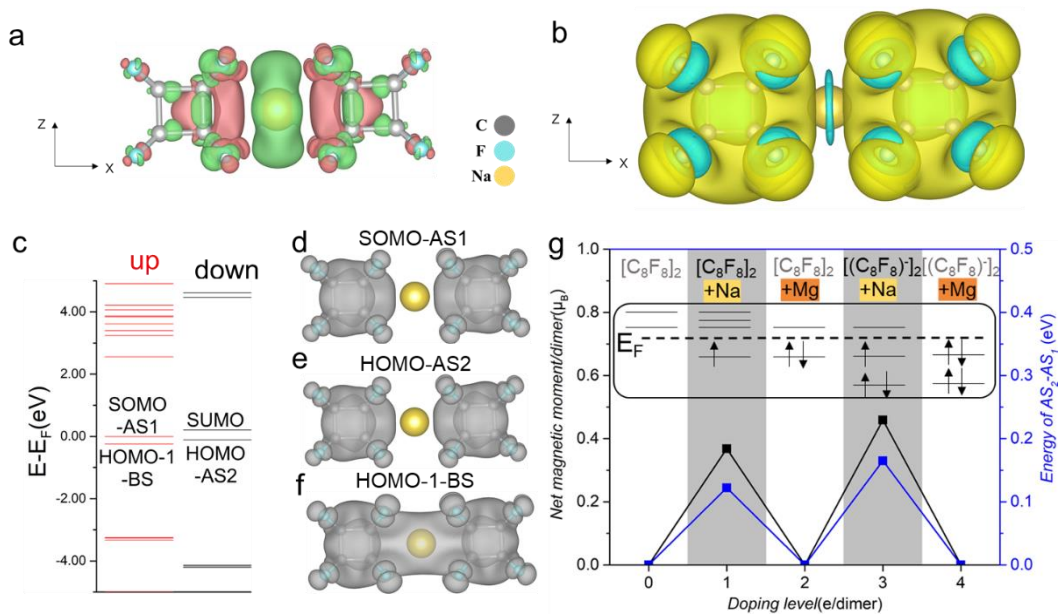

**Fig. 4 Effect of Na/Mg intercalation on $C_8F_8$ dimers.** (a) Inter-molecular differential charge density of Na-doped $[C_8F_8]^-$ dimers. An isosurface value of 0.001 $e$/Bohr$^3$ was used. (b) Spin density of Na-doped $[C_8F_8]^-$ dimers. The isosurface value was set to 3e-5 $e$/Bohr$^3$. (c) Schematic of molecular orbital energy levels. Energy level of singly occupied molecular orbital (SOMO) is set to zero as reference. (d-f) Wavefunction norms for SOMO (d), HOMO (e) and HOMO-1 (f), respectively. The isosurface value was set to 0.0005 $e$/Bohr$^3$. (g) Net magnetic moment and energy difference of $E_{AS1}$-$E_{AS2}$ as a function of electron doping level.

Alkaline earth metal Na and Mg doped 1D $[C_8F_8]^-$ show similar electron-filling number tunable magnetism, thus leading to FM 1D Na-doped $C_8F_8$ chain (1 $e$ per $C_8F_8$) and Mg-doped $[C_8F_8]^-$ chain (3 $e$ per $C_8F_8$). The intermolecular magnetic coupling mechanism in Na-doped $C_8F_8$ chain is similar to that in its dimer, dominated by the FM direct exchange mediated by Na atoms (Fig. 5(c-d)). Figure 5a shows the electronic bandstructure of the 1D Na-doped $C_8F_8$ chain, indirect spin gapless semiconductor feature can be observed with VBM and CBM show opposite spin polarization direction, indicating excellent tunability and potential spintronics applications [39–41].

The band structure of uniform 1D FM Mg-doped $[C_8F_8]^-$ chain is shown in Fig. 5d. Two dispersive bands cross the Fermi level roughly at the point (0.25, 0, 0), indicating possible dimerization due to Peierls instability. By manipulating the positions of Mg atoms and $C_8F_8$ molecules, we found that the dimerization shown in the figure (5f) results in 0.09 eV per $C_8F_8$ lowered total energy and a band gap opening of 0.59 eV (Fig. 5e)., Two $[C_8F_8]^-$ move close to each other by 0.23 Å in the dimerization process. After the band gap opening, the electron filling is closer to the case of Mg-doped $[C_8F_8]^-$ dimer, resulting in the non-magnetic groundstate and a new candidate material for further study of one-dimensional CDW chains.

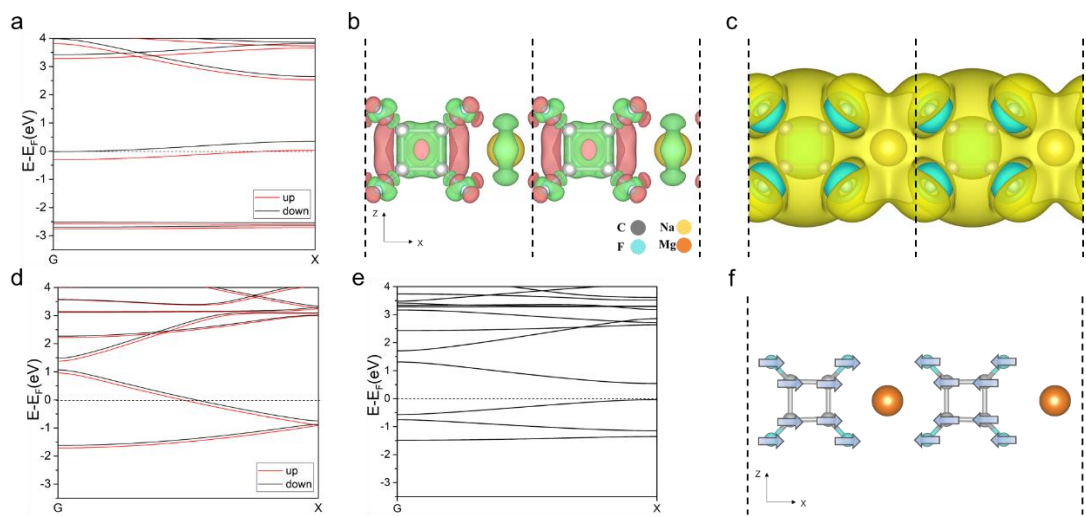

**Fig. 5 Na/Mg-doped $C_8F_8$ 1D chains.** (a) Band structure of a Na-doped $C_8F_8$ chain. (b) Intermolecular differential charge density of a Na-doped $C_8F_8$ chain. An isosurface value of 0.0018 $e$/Bohr$^3$ was used. (c) Spin density of a Na-doped $C_8F_8$ chain. The isosurface value was set to 7e-5 $e$/Bohr$^3$. (d-e) Band structure before (d) and after (e) Peierls distortion of a Mg-doped $[C_8F_8]^-$ chain. (f) Schematic of 1 x 2 CDW structure of a Mg-doped $[C_8F_8]^-$ chain. Gray arrows correspond to the direction of atomic displacement after dimerization.

# IV. CONCLUSIONS

In summary, we theoretically predicted the inter-molecule magnetism and electronic structures are tunable by stacking orders and alkaline earth metal atom intercalation in $C_8F_8$ dimers and 1D chains. By comparing the bonding energies of $[C_8F_8]^-$ dimers with different constructed stacking orders, we managed to find the **d11-c** stacking with the inter-molecular F atoms directly opposite the $C_4$ ring of the other molecule is the most stable with intermolecular AFM coupling. However, FM order becomes the ground state in the most unstable d11 stacking with intermolecular F atoms exactly face the F atoms in the other molecule. We found the inter-molecule couplings are dominated by noncovalent C-F⋯$C_4$ halogen interaction. The relative position of the lone pairs and σ-holes at the molecular interface are tunable by stacking orders, thus modifying the magnetic groundstates of dimers. We extend our calculations to 1D $[C_8F_8]^-$ chains and observed similar noncovalent C-F⋯$C_4$ halogen interaction dominated intermolecular AFM couplings.

We further considered alkaline earth metal Na and Mg intercalation on $C_8F_8$ dimers and 1D chains. Na/Mg intercalations stabilize the originally most unstable d11 stacking and allows the ferromagnetism to become the ground state. Intercalated atoms bond with adjacent $C_4$ rings and form $C_4$-M-$C_4$ bonds, leading to FM direct exchange at the inter-molecule region. The existence of the magnetic moment of dimers relies on the electron-filling number of the four midgap states, which is tunable by substituting intercalated atoms or electron doping. We found the odd-even oscillation of magnetic moment and the energy difference $E_{AS1}$-$E_{AS2}$ as a function of electron doping level, indicating atomic intercalation to be a promising approach manipulating the on/off of magnetic moment of $C_8F_8$ dimer. In addition, we found 1D Na-doped $C_8F_8$ chain exhibits indirect spin gapless semiconductor feature, indicating excellent tunability and potential spintronics applications. Meanwhile, 1D Mg-doped $[C_8F_8]^-$ chain would undergoes dimerization due to Peierls instability and transforms into a NM CDW phase.

The above results manifest the significant roles of stacking order and intercalation in modifying intermolecular magnetism of dimers and 1D chains, which shall boost magnetic applications of cluster-assembled materials of clusters in spintronics and optoelectronics. The revealed halogen bond-dominated exchange mechanisms is paramount additions the previous reported non-covalent interactions, providing a new possibility to regulate the magnetic and electronic properties of low-dimensional materials.

## ACKNOWLEDGEMENTS

We gratefully acknowledge financial support from the Ministry of Science and Technology (MOST) of China (Grant No. 2018YFE0202700), the National Key R&D Program of China (Grant No. 2023YFA1406500), the National Natural Science Foundation of China (Grants No. 11974422, 12104504 and 12204534), the Strategic Priority Research Program of the Chinese Academy of Sciences (Grant No. XDB30000000), the Fundamental Research Funds for the Central Universities, and the Research Funds of Renmin University of China (Grants No. 22XNKJ30). All calculations for this study were performed at the Physics Lab of High-Performance Computing (PLHPC) and the Public Computing Cloud (PCC) of Renmin University of China.

# Reference

[1] P. Jena, S. N. Khanna, and C. Yannouleas, *Comment on '"Patterns and Barriers for Fission of Charged Small Metal Clusters,"'* Phys. Rev. Lett. **69**, 1471 (1992).

[2] S. N. Khanna and P. Jena, *Atomic Clusters: Building Blocks for a Class of Solids*, Phys. Rev. B **51**, 13705 (1995).

[3] Z. Luo and A. W. Castleman, *Special and General Superatoms*, Acc. Chem. Res. **47**, 2931 (2014).

[4] Z. Luo and S. Lin, *Advances in Cluster Superatoms for a 3D Periodic Table of Elements*, Coordination Chemistry Reviews **500**, 215505 (2024).

[5] Y. Jia, C.-Q. Xu, C. Cui, L. Geng, H. Zhang, Y.-Y. Zhang, S. Lin, J. Yao, Z. Luo, and J. Li, *Rh19−: A High-Spin Super-Octahedron Cluster*, SCIENCE ADVANCES (2023).

[6] Y. Jia, J. Li, M. Huang, L. Geng, H. Zhang, S.-B. Cheng, Y. Yi, and Z. Luo, *Ladder Oxygenation of Group VIII Metal Clusters and the Formation of Metalloxocubes M13O8+*, J. Phys. Chem. Lett. **13**, 733 (2022).

[7] Q. Du, Z. Wang, S. Zhou, J. Zhao, and V. Kumar, *Searching for Cluster Lego Blocks for Three-Dimensional and Two-Dimensional Assemblies*, Phys. Rev. Materials **5**, 066001 (2021).

[8] R. Robles and S. N. Khanna, *Magnetism in Assembled and Supported Silicon Endohedral Cages: First-Principles Electronic Structure Calculations*, Phys. Rev. B **80**, 115414 (2009).

[9] H.-G. Xu, X.-Y. Kong, X.-J. Deng, Z.-G. Zhang, and W.-J. Zheng, *Smallest Fullerene-like Silicon Cage Stabilized by a V2 Unit*, The Journal of Chemical Physics **140**, 024308 (2014).

[10] J. Chen, R. Liao, L. Sai, J. Zhao, and X. Wu, *B63: The Most Stable Bilayer Structure with Dual Aromaticity*, J. Phys. Chem. Lett. **15**, 4167 (2024).

[11] W.-L. Li, X. Chen, T. Jian, T.-T. Chen, J. Li, and L.-S. Wang, *From Planar Boron Clusters to Borophenes and Metalloborophenes*, Nat Rev Chem **1**, 0071 (2017).

[12] D. A. Bonhommeau, R. Spezia, and M.-P. Gaigeot, *Charge Localization in Multiply Charged Clusters and Their Electrical Properties: Some Insights into Electrospray Droplets*, The Journal of Chemical Physics **136**, 184503 (2012).

[13] M. A. Miller, D. A. Bonhommeau, C. J. Heard, Y. Shin, R. Spezia, and M.-P. Gaigeot, *Structure and Stability of Charged Clusters*, J. Phys.: Condens. Matter **24**, 284130 (2012).

[14] X. Chen, Y. Li, and X. Zhao, *Screening of Transition Metal-Based MOF as Highly Efficient Bifunctional Electrocatalysts for Oxygen Reduction and Oxygen Evolution*, Surfaces and Interfaces **38**, 102821 (2023).

[15] N. Xia et al., *Assembly-Induced Spin Transfer and Distance-Dependent Spin Coupling in Atomically Precise AgCu Nanoclusters*, Nat Commun **13**, 5934 (2022).

[16] Y. Zhao, Y. Guo, Y. Qi, X. Jiang, Y. Su, and J. Zhao, *Coexistence of Ferroelectricity and Ferromagnetism in Fullerene-Based One-Dimensional Chains*, Advanced Science **10**, 2301265 (2023).

[17] Z. Liu, X. Wang, J. Cai, and H. Zhu, *Room-Temperature Ordered Spin Structures in Cluster-Assembled Single V@Si$_{12}$ Sheets*, J. Phys. Chem. C **119**, 1517 (2015).

[18] C.-I. Yang, Z.-Z. Zhang, and S.-B. Lin, *A Review of Manganese-Based Molecular Magnets and Supramolecular Architectures from Phenolic Oximes*, Coordination Chemistry Reviews **289–290**, 289 (2015).

[19] S. A. Claridge, A. W. Castleman, S. N. Khanna, C. B. Murray, A. Sen, and P. S. Weiss, *Cluster-Assembled Materials*, ACS Nano **3**, 244 (2009).

[20] T. Kato and T. Yamabe, *Electron–Phonon Interactions in Charged Cubic Fluorocarbon Cluster, (CF)8*, The Journal of Chemical Physics **120**, 1006 (2004).

[21] S. M. Spyrou, I. Sauers, and L. G. Christophorou, *Electron Attachment to the Perfluoroalkanes n -C N F2 N +2 ( N =1–6) and i -C4F1 a)*, The Journal of Chemical Physics **78**, 7200 (1983).

[22] G. L. Gutsev and L. Adamowicz, *Relationship between the Dipole Moments and the Electron Affinities for Some Polar Organic Molecules*, Chemical Physics Letters **235**, 377 (1995).

[23] C. S. Wannere, K. W. Sattelmeyer, H. F. Schaefer III, and P. von R. Schleyer, *Aromaticity: The Alternating C C Bond Length Structures of [14]-, [18]-, and [22]Annulene*, Angewandte Chemie International Edition **43**, 4200 (2004).

[24] K. K. Irikura, *Sigma Stellation: A Design Strategy for Electron Boxes*, J. Phys. Chem. A **112**, 983 (2008).

[25] M. Sugiyama, M. Akiyama, Y. Yonezawa, K. Komaguchi, M. Higashi, K. Nozaki, and T. Okazoe, *Electron in a Cube: Synthesis and Characterization of Perfluorocubane as an Electron Acceptor*, Science **377**, 756 (2022).

[26] P. E. Blöchl, *Projector Augmented-Wave Method*, Phys. Rev. B **50**, 17953 (1994).

[27] G. Kresse and D. Joubert, *From Ultrasoft Pseudopotentials to the Projector Augmented-Wave Method*, Phys. Rev. B **59**, 1758 (1999).

[28] G. Kresse and J. Furthmüller, *Efficient Iterative Schemes for Ab Initio Total-Energy Calculations Using a Plane-Wave Basis Set*, Phys. Rev. B **54**, 11169 (1996).

[29] S. Grimme, J. Antony, S. Ehrlich, and H. Krieg, *A Consistent and Accurate Ab Initio Parametrization of Density Functional Dispersion Correction (DFT-D) for the 94 Elements H-Pu*, The Journal of Chemical Physics **132**, 154104 (2010).

[30] R. Tonner, P. Rosenow, and P. Jakob, *Molecular Structure and Vibrations of NTCDA Monolayers on Ag(111) from Density-Functional Theory and Infrared Absorption Spectroscopy*, Phys. Chem. Chem. Phys. **18**, 6316 (2016).

[31] M. Stein and M. Heimsaat, *Intermolecular Interactions in Molecular Organic Crystals upon Relaxation of Lattice Parameters*, Crystals **9**, 665 (2019).

[32] P. Jiang, C. Wang, D. Chen, Z. Zhong, Z. Yuan, Z.-Y. Lu, and W. Ji, *Stacking Tunable Interlayer Magnetism in Bilayer CrI 3*, Phys. Rev. B **99**, 144401 (2019).

[33] P. Li, C. Wang, J. Zhang, S. Chen, D. Guo, W. Ji, and D. Zhong, *Single-Layer CrI3 Grown by Molecular Beam Epitaxy*, Science Bulletin **65**, 1064 (2020).

[34] D. B. Dougherty, M. Feng, H. Petek, J. T. Yates, and J. Zhao, *Band Formation in a Molecular Quantum Well via 2D Superatom Orbital Interactions*, Phys. Rev. Lett. **109**, 266802 (2012).

[35] J. Zhao, M. Feng, D. B. Dougherty, H. Sun, and H. Petek, *Molecular Electronic Level Alignment at Weakly Coupled Organic Film/Metal Interfaces*, ACS Nano **8**, 10988 (2014).

[36] Z. Han, G. Czap, C. Chiang, C. Xu, P. J. Wagner, X. Wei, Y. Zhang, R. Wu, and W. Ho, *Imaging the Halogen Bond in Self-Assembled Halogenbenzenes on Silver*, Science **358**, 206 (2017).

[37] A. Bauzá, T. J. Mooibroek, and A. Frontera, *Tetrel-Bonding Interaction: Rediscovered Supramolecular Force?*, Angew Chem Int Ed **52**, 12317 (2013).

[38] J. Zhao, Q. Du, S. Zhou, and V. Kumar, *Endohedrally Doped Cage Clusters*, Chem. Rev. **120**, 9021 (2020).

[39] X. L. Wang, *Proposal for a New Class of Materials: Spin Gapless Semiconductors*, Phys. Rev. Lett. **100**, 156404 (2008).

[40] X.-L. Wang, S. X. Dou, and C. Zhang, *Zero-Gap Materials for Future Spintronics, Electronics*

*and Optics*, NPG Asia Mater **2**, 31 (2010).

[41] S. Ouardi, G. H. Fecher, C. Felser, and J. Kübler, *Realization of Spin Gapless Semiconductors: The Heusler Compound Mn 2 CoAl*, Phys. Rev. Lett. **110**, 100401 (2013).

# APPENDIX : SUPPLEMENTED INFORMATION

**Table S1. Detailed distances and binding energies for each configuration.**

| Stacking | distance/A | Eb of FM/eV | Eb of AFM/eV |
|---|---|---|---|
| **d13** | 2.56 | -0.128 | -0.129 |
| **d14-c** | 2.73 | 0.000 | -0.161 |
| **d11-c** | 2.79 | -0.091 | -0.202 |
| **d15-c** | 3.09 | 0.085 | -0.118 |
| **d12-c** | 3.26 | -0.070 | -0.134 |
| **d14** | 4.08 | -0.059 | -0.077 |
| **d12** | 4.14 | -0.058 | -0.126 |
| **d15** | 4.44 | -0.036 | -0.067 |
| **d44** | 5.48 | 0.013 | 0.013 |
| **d33** | 5.84 | 0.019 | 0.019 |
| **d11-c** | 6.80 | 0.011 | 0.034 |

**Table S2. Detailed binding energy of d14 after translation.**

| Configuration | Mag. Config. | Eb(eV) |
|---|---|---|
| **d14** | FM | -0.06 |
|  | AFM | -0.08 |
| **d14-c** | FM | 0.00 |
|  | AFM | -0.16 |
| **t-C1** | FM→AFM | -0.10 |
|  | AFM |  |
| **t-C2** | FM | 0.02 |
|  | AFM | 0.00 |
| **t-CC1** | FM | -0.10 |
|  | AFM | -0.10 |
| **t-CC2** | FM | -0.10 |
|  | AFM | -0.10 |
| **t-CC3** | FM | -0.10 |
|  | AFM | -0.10 |

**Table S3. Detailed binding energy of d12 after translation.**

| Stacking | Mag. Config. | Eb(eV) |
|---|---|---|
| **d12** | FM | -0.06 |
|  | AFM | -0.13 |
| **d12-c** | FM | -0.07 |
|  | AFM | -0.13 |
| **t-C1** | AFM | -0.06 |
| **t-C2** | AFM | -0.11 |
| **t-CC1** | AFM | -0.05 |
| **t-CC2** | AFM | -0.07 |
| **t-CC3** | AFM | -0.09 |

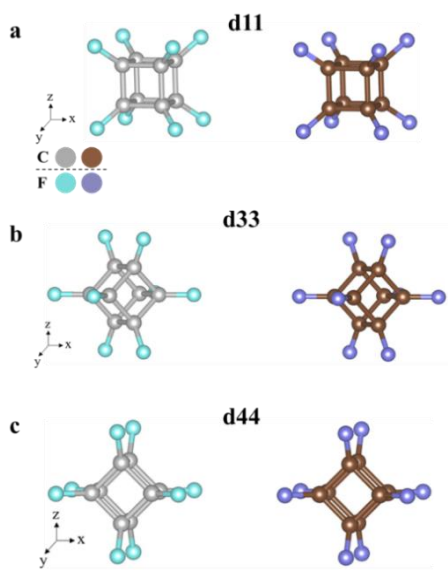

**Figure S1.** The structures after relaxation of the three AA stacking configurations.

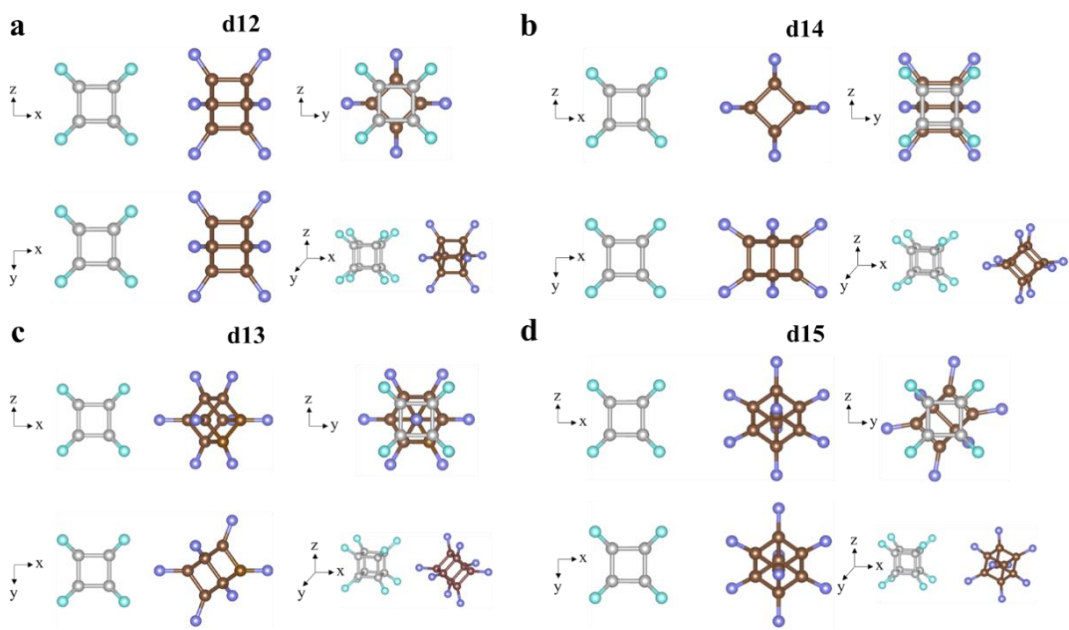

**Figure S2.** The structure after relaxation of four AB stacking configurations after considering rotation.

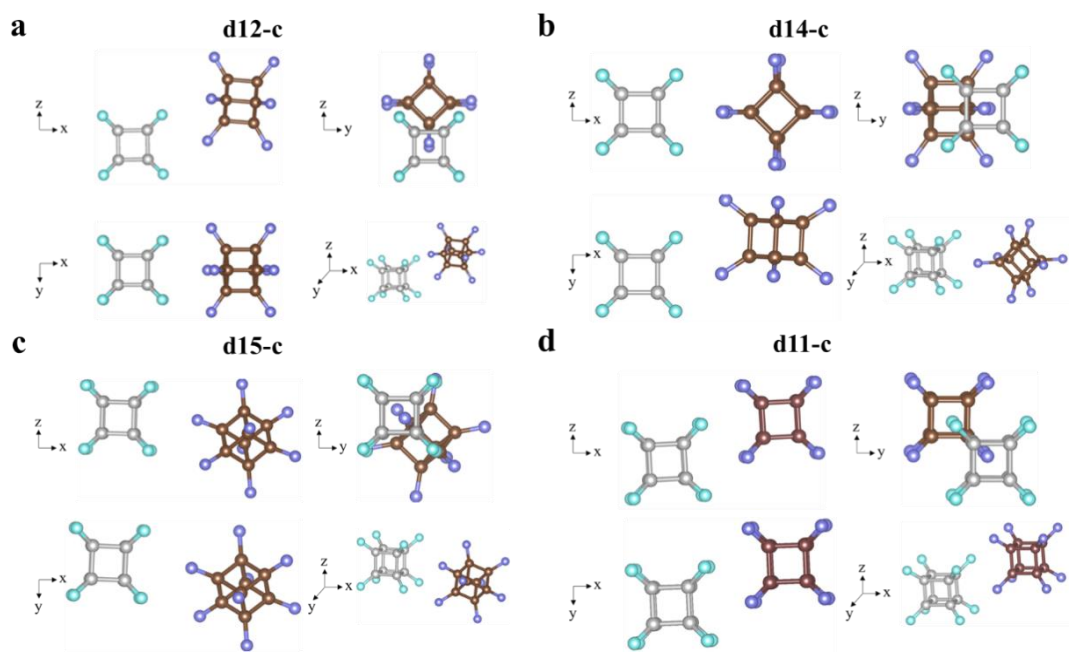

**Figure S3.** The structure after relaxation of four configurations after rotation and translation operations.

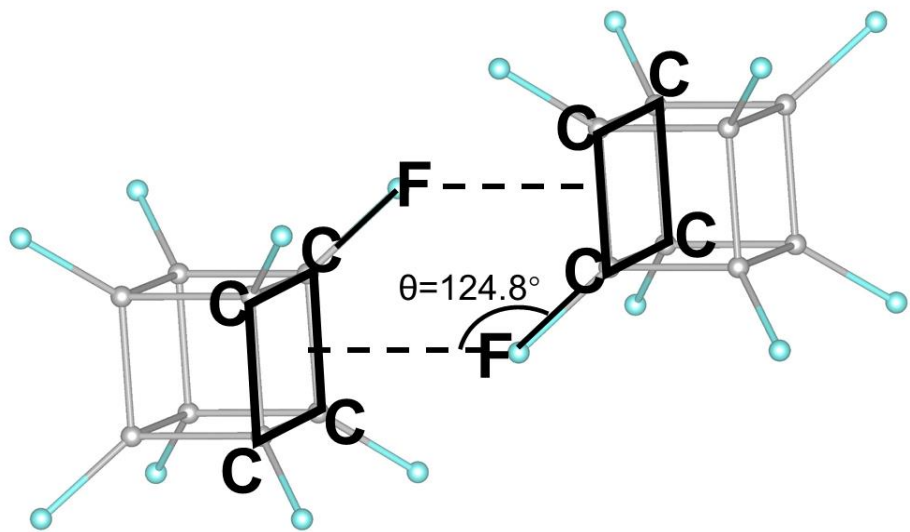

**Figure S4. Schematic representation of the halogen-like bonds in the $C_8F_8$ dimer.**

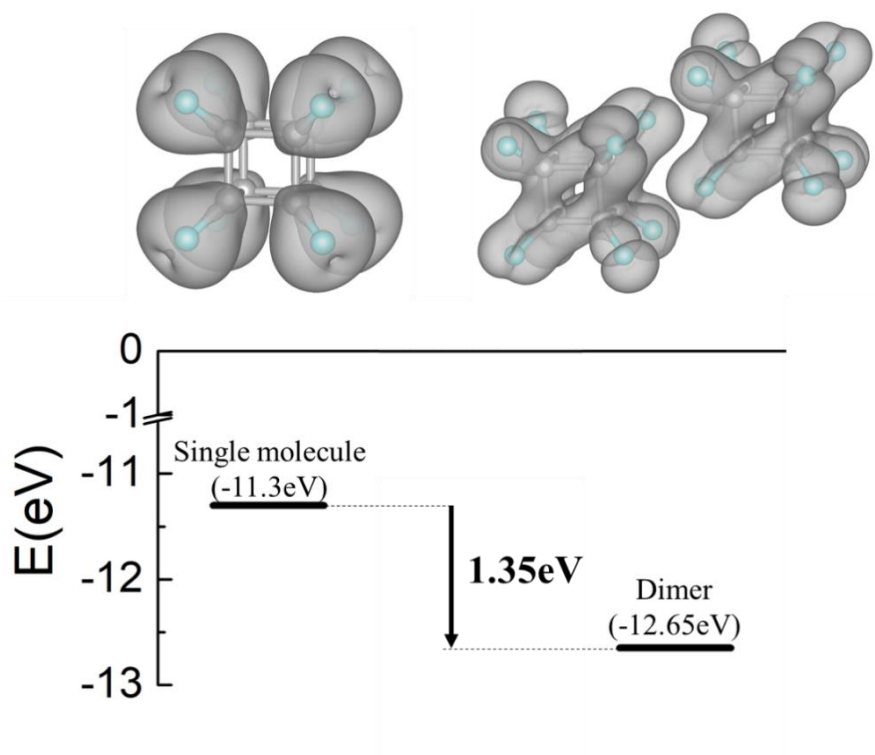

**Figure S5. The C-F bond energy in single molecule and dimer.** The isosurface value of corresponding wavefunction norms was set to 0.005 $e$/Bohr$^3$.

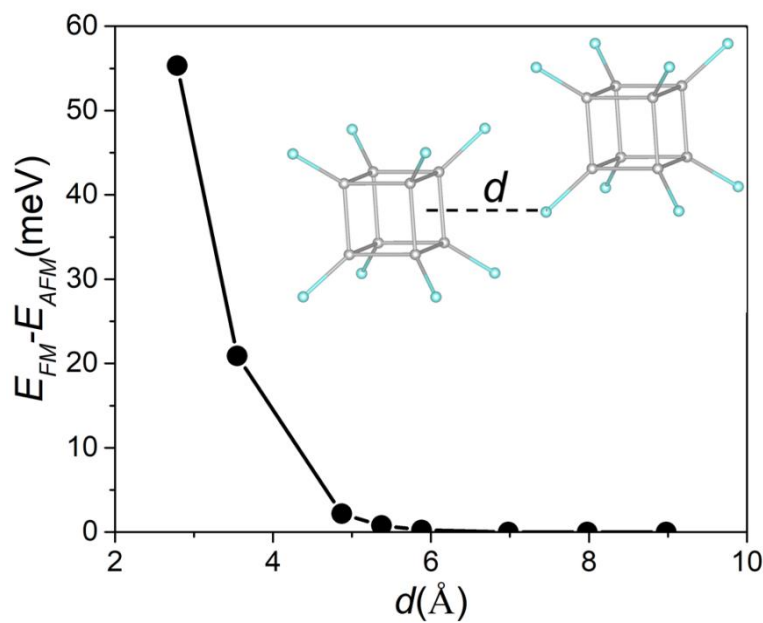

**Figure S6. A plot of the ferromagnetic-antiferromagnetic energy difference $E_{FM}$-$E_{AFM}$ versus the molecular spacing $d$.** Here $d$ denotes the distance from the F atom to the center of the cyclobutane ring of another cluster, as shown in the upper right corner of the figure.

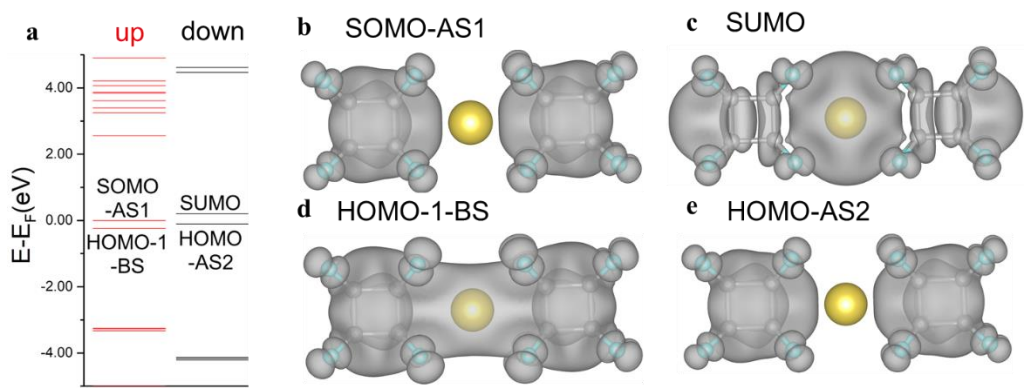

**Figure S7. Detailed wavefunction norms of Na-doped [$C_8F_8$]$^-$ dimer.** (a) Schematic of molecular orbital energy levels. Energy level of singly occupied molecular orbital (SOMO) is set to zero as reference. (b-e) Wavefunction norms for SOMO (b), SUMO (c), HOMO-1 (d) and HOMO (e), respectively. The isosurface value was set to 0.0005 $e$/Bohr$^3$.

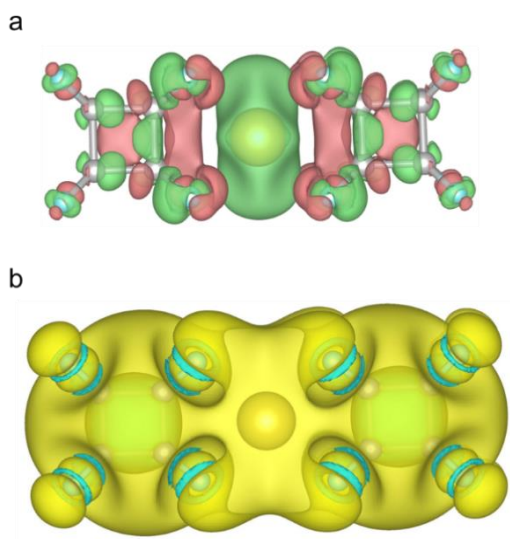

**Figure S8. Na-doped C$_8$F$_8$ dimer.** (a) Inter-molecular differential charge density of Na-doped C$_8$F$_8$ dimers. An isosurface value of 0.001 $e$/Bohr$^3$ was used. (b) Spin density of Na-doped C$_8$F$_8$ dimers. The isosurface value was set to 8e-5 $e$/Bohr$^3$.